\documentclass[sigconf]{acmart}
\acmConference[XXX Coference]{XXX}{Time}{City}
\renewcommand\footnotetextcopyrightpermission[1]{} 
\usepackage{textcomp}
\usepackage{enumitem}
\usepackage{graphicx}
\usepackage{textcomp}
\usepackage{algpseudocode}
\usepackage{subfigure}
\usepackage{xcolor}
\usepackage{enumitem}
\usepackage{diagbox}
\usepackage{amsfonts}
\usepackage{amsmath}
\usepackage{amsthm}
\usepackage{multirow}
\usepackage{array}
\usepackage{times}
\usepackage{epsfig}
\usepackage{listings}
\usepackage{bm}
\newcommand{\tabincell}[2]{\begin{tabular}{@{}#1@{}}#2\end{tabular}}
\begin{document}
\title{A Hybrid Approach to Fine-grained Automated Fault Localization}
\author{Leping Li}
\email{3120170473@bit.edu.cn}
\author{Hui Liu}
\email{liuhui08@bit.edu.cn}

\begin{abstract}

Fault localization is to identify faulty source code. It could be done on various granularities, e.g., classes, methods, and statements. Most of the automated fault localization (AFL) approaches are coarse-grained because it is challenging to accurately locate fine-grained faulty software elements, e.g., statements. However, some automated approaches, e.g., automated program repair (APR), require fine-grained fault localization. As a result, such APR approaches often leverage the traditional spectrum-based fault localization (SBFL) techniques although more advanced coarse-grained AFL approaches have been proposed recently. SBFL, based on dynamic execution of test cases only, is simple, intuitive, and generic (working on various granularities). However, its accuracy deserves significant improvement. To this end, in this paper, we propose a hybrid  fine-grained AFL approach  based on both dynamic spectrums and static statement types. The rationale of the approach is that some types of statements are significantly more/less error-prone than others, and thus statement types could be exploited for fault localization.  On a crop of faulty programs, we compute the error-proneness for each type of statements, and assign priorities to special statement types that are steadily more/less error-prone than others. For a given faulty program under test, we first leverage traditional spectrum-based fault localization algorithm to identify all suspicious statements and to compute their  suspicious scores. For each of the resulting suspicious statements, we retrieve its statement type as well as the special priority associated with the type. The final suspicious score  is the product of the SBFL suspicious score and the priority assigned to the statement type.  A significant advantage of the approach is that it is simple and intuitive, making it efficient and easy to interpret/implement. We evaluate the proposed approach on widely used benchmark Defects4J. The evaluation results suggest that the proposed approach outperforms widely used SBFL, reducing the absolute waste effort (AWE)  by 9.3\% on average.

\end{abstract}



\maketitle

\section{Introduction}
\label{section:Introduction}
Fault localization is to identify faulty source code in programs~\cite{APRSurvey,Purification,FIFL}. Accurate fault localization is critical for bug fixing. Before developers can fix faulty programs, they have to accurately identify faulty source code from the programs. However, manual fault localization is challenging, tedious, and time-consuming~\cite{Economic}, especially for inexperienced developers in debugging complex programs developed by others. To this end, a few automated fault localization techniques (called AFL for short) have been proposed~\cite{AAAI20,BiDireAttention,Scaffle}. Such novel approaches have significantly reduced the cost of fault localization, thus also reduced the cost of software debugging that often takes up to 80\% of the total software cost~\cite{Economic}.

Fault localization could be done on various granularities, e.g., classes~\cite{IRICPC20}, methods~\cite{DEEPFL,ProFL}, and statements~\cite{DStar}. Most of the automated fault localization approaches  are coarse-grained~\cite{PREDFL} because it is challenging to automatically and accurately locate fine-grained faulty software elements, e.g., statements. For example, with the widely used spectrum-based fault localization techniques (called SBFL for short), developers may have to inspect hundreds of statements manually before a faulty statement is finally identified~\cite{DStar}. The effort wasted in inspecting such bug-free statements is called \emph{absolute wasted effort} (called AWE for short)~\cite{PageRank}.

However, some automated approaches, e.g., automated program repair~\cite{ACS,ssfix,Angelix,PAR,CapGen,GenProg} (called APR for short) requires fine-grained fault localization~\cite{APRSurvey}. To the best of our knowledge, most of the ARP tools require statement-level/line-level fault localization. For example, mutation-based ARP tools like PraPR~\cite{PraPR}  generate candidate patches by applying predefined mutation operators to  suspicious statements that are identified by fine-grained fault localization approaches. SimFix~\cite{SimFix} requires line-level fault localization to extract the contexts of the faulty source code, and generates candidate patches based on the extracted contexts.  As a result, such APR algorithms/tools often leverage the traditional spectrum-based fault localization algorithms although more advanced coarse-grained AFL approaches have been proposed recently~\cite{DEEPFL,IRICPC20}. Spectrum-based fault localization, based on dynamic execution of test cases only, is simple, intuitive, and generic (working on various granularities). However, its accuracy deserves significant improvement~\cite{ProFL,DEEPFL}. If fine-grained fault localization is inaccurate, automated program repair tools have to generate and validate numerous candidate patches on a large number of bug-free statements before they finally come to the faulty statement.  As a result, cost of such APR tools could be unnecessarily high.

To this end, in this paper, we propose an automated, simple, and intuitive hybrid approach (called \emph{HybridAFL}) to fine-grained fault localization. The rationale of the approach is that some types of statements are significantly more/less error-prone than others, and thus the statement types could be exploited for fault localization. Further more, such static type information may complement dynamic spectrums of test case executions~\cite{Tarantula}. Consequently, the proposed approach exploits both dynamic spectrums of test case executions and static statement types. On a crop of faulty programs, we leverage simple statistics to compute the error-proneness for each type of statements, and assign  priorities to special statement types that are steadily more/less error-prone than others. For a given faulty program under test, we first leverage traditional spectrum-based fault localization algorithm to identify all suspicious statements and to compute their  suspicious scores. For each of the resulting suspicious statements, we retrieve its statement type as well as the special priority associated with the type. The final suspicious score  is the product of the SBFL suspicious score and the priority assigned to the statement type.  We evaluate the proposed approach on widely used benchmark Defects4J~\cite{D4j}. Our evaluation results suggest that the proposed approach significantly outperforms the widely used SBFL techniques. It significantly reduces the absolute wasted effort (AWE) on five out of the seven programs, with an average reduction of 9.3\% on all of the involved faulty programs.

The paper makes the following contributions:
\begin{itemize}[leftmargin=0.5cm,topsep=0.06cm]
  \item A simple, intuitive, but accurate approach to fine-grained fault localization. To the best of our knowledge, it is the first statement-level fault localization approach that leverages both dynamic and static information of the faulty programs. A significant advantage of the approach is that it is simple and intuitive, making it reliable, efficient,  and easy to interpret/implement.
  \item An extensive evaluation of the proposed approach on well-known dataset (Defects4J). The evaluation results suggest that the proposed approach can significantly improve the state of the art in fine-grained fault localization.
\end{itemize}

\section{Related Work}
\label{section:RelatedWork}
\subsection{Fine-grained Fault Localization}
Spectrum-based fault localization (SBFL)~\cite{Tarantula,Ochiai,Ochiai2} is one of the most widely used techniques that could be employed to locate both fine-grained and coarse-grained faulty source code~\cite{AFLSurvey,APRSurvey}.
Spectrums refer to the program execution traces of successful and failed executions, i.e., run-time profiles about which program entities are covered by each test.  The key idea of spectrum-based fault localization is that a program entity covered by more failing tests but less passing tests is more likely to be faulty. To this end, SBFL techniques run test cases and generate spectrums based on the executions. Based on the resulting spectrums, SBFL techniques can leverage different formulae to estimate the suspicious scores for all software entities covered by failed executions. For example,
Tarantula~\cite{Tarantula} computes suspicious scores as follows:
\begin{equation}\label{eq:SBFLT}
sus(e) = \frac{\frac{CF}{CF+UF}}{\frac{CF}{CF+UF} + \frac{CP}{CP+UP}}
\end{equation}
where $sus(e)$ is the suspicious score of element $e$, $CP$ is the number of passed tests covering element $e$, $CF$ is the number of failed tests covering element $e$, $UP$ is the number of passed tests that do not cover  $e$,  $UF$ is the number of failed tests that do not cover  $e$. 
Chen et.al~\cite{Jaccard} compute the suspicious score based on $Jaccard$ similarity as follows:
\begin{equation}\label{eq:SBFLJ}
sus(e) = \frac{CF}{CF+CP+UF}
\end{equation}
Abreu et.al~\cite{Ochiai} compute the suspicious score as follows:
\begin{equation}\label{eq:SBFLO}
sus(e) = \frac{CF}{\sqrt{(CF+UF)*(CF+CP)}}
\end{equation}
Wong et.al~\cite{DStar} compute the suspicious score as follows:
\begin{equation}\label{eq:SBFLD}
sus(e) = \frac{CF^*}{UF+CP}
\end{equation}
where * is a parameter whose value should be no less than 1.

Mutation-based fault localization (MBFL)~\cite{Metallaxis1,Metallaxis2, MUSE} is another way to both fine-grained and coarse-grained fault localization. To the best of our knowledge, the approach proposed by Papadakis and Traon~\cite{MBFL2012} is the first MBFL technique. It predefines mutation operators to generate mutants on specific code elements within a faulty program. It then runs test cases on the generated mutants, and utilizes Ochiai~\cite{Ochiai} formula to compute suspicious scores based on the executions of test cases.  The rationale of MBFL is that mutation on faulty code elements is more likely to make passed tests fail while mutation on correct elements is more likely to make failed tests pass.


State-based fault localization~\cite{VSwitch,PreSwitch} identifies suspicious statements by intentionally updating states of the faulty system while executing failed test cases. For example, during the execution of a failed test case, delta debugging~\cite{VSwitch} replaces the value of a variable with its corresponding value in passed test cases. If the replacement leads to the same fault, the variable is likely bug-free. Otherwise, the variable is suspicious. Predicate switching~\cite{PreSwitch,PathGen} is another well-known state-based fault localization technique. It switches a predicate executed by  failed test cases. If the switch passes the failed test cases,  the modified predicate is suspicious. 

The fault localization approach proposed by Liblit et al.~\cite{Liblit05} computes suspiciousness of a predicate $p$ by estimating two conditional probabilities: the rate of $p$ being  covered  implies failed execution (noted as $P_1$), and the rate of $p$ being true implies failed execution (noted as $p_2$).
If $p_2$ is significantly greater than $p_1$, the predicate $p$ is likely to be faulty.
Similarly, SOBER~\cite{SOBER} computes suspiciousness of a predicate by calculating times of the predicate being true divided by times of the predicate being covered with regard to the execution of passed/failed test cases:
\begin{eqnarray}
\pi(p)_{pass}&=&\frac{CP_{true}}{CP_{true}+CP_{false}}\\
\pi(p)_{fail}&=&\frac{CF_{true}}{CF_{true}+CF_{false}}
\end{eqnarray}
where $CF$ is the number of failed test cases covering predicate $p$, and $CP$ is the number of passed test cases covering predicate $p$. Subscripts $true$ and $false$ represent the value of $p$ when associated test cases are executed.  If  $\pi(p)_{pass}$ and $\pi(p)_{fail}$ are significantly different, the predicate $p$ is considered to be suspicious. CBT~\cite{CBT} extends such approach to cover additional entities, e.g., statements and methods.


Our approach is similar to such approaches in that all of them could be employed for fine-grained fault localization. Our approach differs from them in that our approach leverages both dynamic test case execution and static statement types of the source code whereas none of such related approaches leverages any static information of the source code, e.g., statement types.
\begin{figure*}
	\begin{center}
		\includegraphics[width=0.95\textwidth]{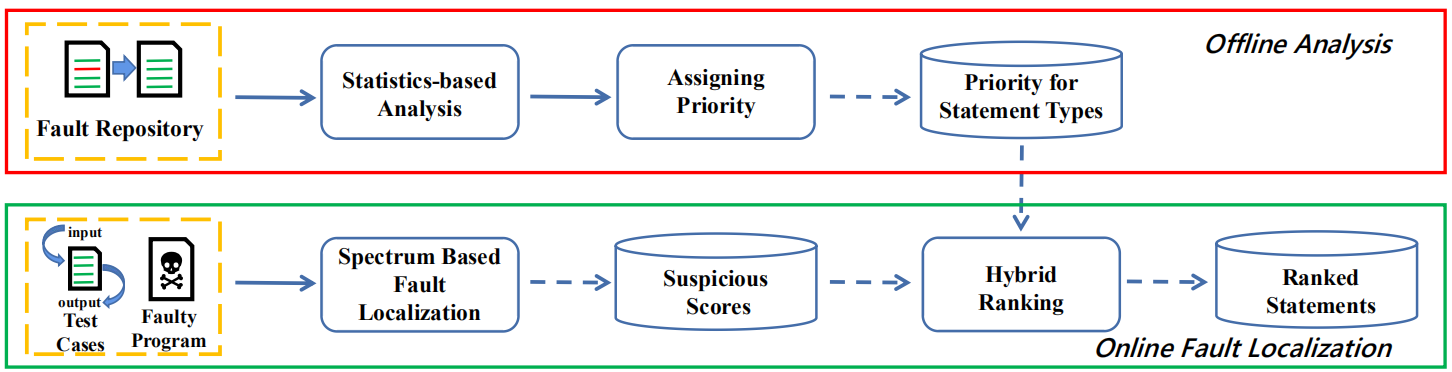}
		\caption{Overview of HybridAFL}
		\label{fig:overview}
	\end{center}
\end{figure*}
\subsection{Coarse-grained Fault Localization}
Coarse-grained fault location is to identify coarse-grained suspicious software entities, e.g., methods and classes. As mentioned in the preceding section,  spectrum-based fault localization (SBFL)~\cite{Ochiai,Ample} and mutation-based fault localization (MBFL)~\cite{MBFL2012,Metallaxis1,MUSE,FIFL} work on both coarse-grained and fine-grained software entities, varying from statements to modules. Besides such generic fault localization approaches, some approaches are specially designed for coarse-grained fault localization. For example, information retrieval (IR) based fault location~\cite{IR1,IR2} ranks potentially faulty elements (methods or classes) by computing their lexical similarity with bug reports. BLUiR~\cite{BLUiR}, proposed by R.K.Saha, parses source file and extracts identifiers as well as code structure information. Based on such information, it leverages IR techniques to compute the similarity between source code and bug reports. 

Learn-to-rank  models  are recently employed for fault location~\cite{TraPT,DEEPFL,IEICE2017}. TraPT~\cite{TraPT} leverages both SBFL and MBFL scores, and employs SVM to learn how to rank suspicious software entities by SBFL and MBFL scores.  DeepFL~\cite{DEEPFL} leverages both dynamic information (SBFL and MBFL scores) and static information, e.g., code complexity and the text of failed test cases. Based on such hybrid information, DeepFL employs deep learning techniques to learn how to rank suspicious software entities.

ProFL~\cite{ProFL} is the first to leverage automated program repair (APR) for fault localization. It employs a mutation-based APR technique, called PraPR~\cite{PraPR}, to generate patches for each program element. Among such patches, it identifies high quality patches that pass more test cases. Software entities that are mutated to create high quality patches are considered as more suspicious entities.

\subsection{Empirical Study on Fault Localization}
Qi et.al~\cite{GenProgFL} conducted an empirical study on the effectiveness of various fault localization techniques with regard to automated program repair. Their results suggest that  $Jaccard$~\cite{Jaccard} frequently outperform other SBFL variants. The results also suggest that accurate fault localization could reduce the number of  candidate patches generated and validated by ARP tools, and thus improves their efficiency.
Yang et.al~\cite{ICSME2017} conducted an empirical study and found that suspiciousness-first techniques outperform rank-first techniques in both patch diversity and parallelity of program repair. Tung et.al~\cite{IRICPC2017} conducted an empirical study on enhancing IR-based fault localization techniques with coverage and code slicing information.  Their results suggest that both coverage and code slicing information could improve the performance of IR-based fault localization techniques.
S.Khatiwada et.al~\cite{IRICPC20} conducted an empirical study whose results suggest that combining these orthogonal IR techniques  targeting different aspects of similarity could enhance the confidence in results and helps to retrieve links that are missed by individual techniques. 
Jiang et.al~\cite{PREDFL} conducted a systematic analysis on different combinations of spectrum-based fault localization techniques and dynamic debugging techniques. Their results suggest that the combinations are often effective.

One of the common findings of such empirical studies is that hybrid approaches leveraging various orthogonal but complementary information are promising. Our approach is in line with this finding. To the best of our knowledge, it is the first hybrid approach to statement-level fault localization, leveraging both test case executions and static statement types.  

\section{Approach}
\label{section:Approach}

\subsection{Overview}
\label{sub:Overview}
Figure~\ref{fig:overview} presents the  overview of the proposed approach (HybridAFL). It is composed of two parts: offline analysis and online fault localization. The former analyzes large defect repositories, identifies significantly more/less error-prone statement types, and assigns special priorities to them. Based on the resulting priorities, the latter collects suspicious statements in a faulty program, computes suspicious scores for each of them, and ranks such statements. Overall, HybridAFL works as follows:
\begin{itemize}[leftmargin=0.5cm,topsep=0.06cm]
\item HybridAFL applies statistics analysis  to a fault repository to compute the error-proneness for each of the statement types involved in the repository.
\item Based on the resulting error-proneness, HybridAFL identifies special statement types that are steadily more/less error-prone than others.
\item HybridAFL assigns priorities to special statement types selected in the preceding step. The more error-prone  a statement  type is, the greater priority it receives.
\item For a faulty program to be debugged, HybridAFL collects  suspicious statements from the faulty program  and computes their suspicious scores by traditional SBFL.
\item HybridAFL adjusts the suspicious scores according to statement types, i.e., multiplying the SBFL suspicious scores by the priorities assigned to the statement types.
\item All of the suspicious statements are ranked  by the updated suspicious scores in descending order. Statements on the top are more likely to be faulty and thus should be inspected first.
\end{itemize}
Notably, the offline analysis could be done once an for all whereas the online fault localization should be conducted independently on each of the faulty programs where faulty statements should be located.
Details of the key steps are presented in the following sections.

\subsection{Statistics-based Analysis on Fault Repositories}
\label{sub:Statistics}
The key rationale of the proposed approach is that different types of statements are not equally error-prone. Some of them could be significantly more error-prone than others, and thus they should be given greater priorities during fault localization. To allocate quantitative priorities to different statement types, we apply statistics-based analysis to a given fault repository (noted as $rep$) containing a set of faulty programs:
\begin{eqnarray}
rep=\{pr_1, pr_2\dots, pr_n\}
\end{eqnarray} where $pr_i$ is the $i$-th faulty program, and $n$ is the size of the repository.  Each program $pr_i\in rep$ is associated with a set of faulty versions.
Each version of the faulty program is accompanied by a concise patch and one or more triggering test cases. Based on the triggering test cases, we collect all suspicious statements in $k$-th version ($V_{i,k}$) of program $pr_i$: Statements covered by any of the triggering test cases are suspicious.  The resulting suspicious statements are noted as $SS(V_{i,k})$. We can also identify all faulty statements in the faulty version, i.e., all statements deleted or modified by the associated patch. The resulting faulty statements are noted as $FS(V_{i,k})$. All suspicious statements associated with  program $pr_i$ are noted as $TSS(pr_i)$:
\begin{equation}\label{eq:suspiciousStatement}
TSS(pr_i)=\bigcup_{k=1}^{TV(pr_i)}{SS(V_{i,k})}
\end{equation}
where $TV(pr_i)$ is the total number of faulty versions of program $pr_i$.
All faulty statements associated with program $pr_i$ are noted as $TFS(pr_i)$:
\begin{equation}
TFS(pr_i)=\bigcup_{k=1}^{TV(pr_i)}{FS(V_{i,k})}
\end{equation}
The average possibility for a suspicious statement to be faulty is:
\begin{equation}\label{eq:avgEP}
AP(pr_i)=\frac{|TFS(pr_i)|}{|TSS(pr_i)|}
\end{equation}

We further divide the suspicious/faulty statements according to their statement types, i.e., all suspicious/faulty statements of the same type are classified into the same category:
\begin{eqnarray}
TSS(pr_i)&=&\bigcup_{k=1}^{N} TSS(pr_i,t_j)\\
TFS(pr_i)&=&\bigcup_{k=1}^{N} TFS(pr_i,t_j)
\end{eqnarray}
where $N$ is the total number of statement types. $TSS(pr_i,t_j)$ and $TFS(pr_i,t_j)$ are the suspicious/faulty statements of type $t_j$. For a suspicious statement of type $t_j$, its possibility of being faulty is noted as $AP(pr_i, t_j)$:
 \begin{equation}\label{eq:avgPossibility-type}
AP(pr_i, t_j)=\frac{|TFS(pr_i, t_j)|}{|TSS(pr_i, t_j)|}
\end{equation}

\subsection{Assigning Priorities to Special Statements}
\label{sub:priorities}
The proposed approach is built on the assumption that some types of suspicious statements are significantly more/less error-prone than others. In this section, we specify how we identify such statements, and how we assign priorities to them in ranking suspicious statements.

For each type of the suspicious statements in program $pr_i$, we define its \emph{relative error-proneness} as follows:
\begin{equation}\label{eq:EPscore}
	RP(pr_i,t_j)=AP(pr_i, t_j)/AP(pr_i)
\end{equation}
where $AP(pr_i)$ (as defined in Equation~\ref{eq:avgEP}) is the average error-proneness for suspicious statements in program $pr_i$, and $AP(pr_i, t_j)$ (as defined in Equation~\ref{eq:avgPossibility-type}) is the error-proneness for statement type $t_j$.
If $RP(pr_i,t_j)$ is significantly greater/smaller than 1, statements of type $t_j$ are significantly more/less error-prone than others, and thus should be given different priorities in fault localization. For a given statement type $t_j$, we validate whether it deserves special priority as follows:
\begin{enumerate}
\item First, we compute its relative error-proneness over programs as:
\begin{equation}
RP(t_j)=\{RP(pr_1,t_j),RP(pr_2,t_j), \dots , RP(pr_n,t_j)\}
\end{equation}
where $n$ is the number of programs.
\item Second, we normalize $RP(t_j)$ as follows:
\begin{equation}
nor\_RP(t_j)=\{lg(RP(pr_1,t_j)), \dots , lg(RP(pr_n,t_j))\}
\end{equation}
The normalization is applied to balance the magnification factors and minification factors. For example,  $RP(pr_1,t_j)=2$ suggests that the error-proneness of the statement type is twice of the average, and $RP(pr_2,t_j)=0.5$ suggests that the error-proneness of the statement type is half of the average. In both cases, the magnification/minification factor of the relative error-proneness is 2. However, the absolute values of the relative error-proneness (i.e., 2 and 0.5 for the preceding example) are significantly different. To this end, we apply the preprocess, and the resulting values ($lg(2)=0.3$ and $lg(0.5)=-0.3$) share the same absolute value.
\item Statement type $t_j$ deserves a special priority if its relative error-proneness $nor\_RP(t_j)$ is steadily greater/smaller than the average.     As a quantitative condition, statement type $t_j$ deserves a special priority if and only if at least 95\% of the elements in $nor\_RP(t_j)$ are on the same side of zero (greater or smaller than zero). In other words, for a new faulty program, we can guarantee that statements of type $t_j$ on this program is more/less error-prone than the average with a high confidence coefficent of 95\%.  
\item If $t_j$ deserves special priority, we assign the priority  as follows:
\begin{eqnarray}
\label{qua:Wt}
W(t_j) &=& lg^{-1}{avg(nor\_RP(t_j))} \nonumber\\
&&=10^{avg(nor\_RP(t_j))}
\end{eqnarray}
where
\begin{eqnarray}
  avg(nor\_RP(t_j)) &=&  \frac{1}{n}\sum_{i=1}^n{lg(RP(pr_i,t_j))}
\end{eqnarray}
is the average of $nor\_RP(t_j)$, and $lg^{-1}$ is the inverse function of $lg$.
\end{enumerate}
$W(t_j)$ is greater than 1 if $nor\_RP(t_j)$ is steadily greater than the average, and $W(t_j)$ is between 0 and 1 if $nor\_RP(t_j)$ is steadily smaller than the average.
Notably, for statement types that do not deserve special priorities, we assign them the default priority, i.e., 1.

%

\subsection{Hybrid Ranking}
The proposed approach is hybrid in that it leverages traditional SBFL algorithms to compute the initial suspicious scores and then adjusts such scores according to statement types. For a faulty program $p'$ that are accompanied with passed test cases and failed test cases, we collect all suspicious statements that are covered by at least one failed test case. Such suspicious statements are noted as $TSS(p')$ (as defined in Equation~\ref{eq:suspiciousStatement}). For each of the suspicious statements (noted as $ss_j$), we leverage $Ochiai$~\cite{Ochiai} to compute its initial suspicious score (noted as $sus(ss_j)$) according to Equation~\ref{eq:SBFLO}.
Ochiai is selected because it is one of the most widely used spectrum-based fault localization (SBFL) algorithms, and existing studies~\cite{APRSurvey} suggest that it frequently outperforms other SBFL variants.


We further adjust the SBFL scores according to statement types as follows:
\begin{equation}\label{eq:NewScore}
 sus'(ss_j) = sus(ss_j) \times W(T(ss_j))
\end{equation}
where $T(ss_j)$ is the type of statement $ss_j$, and $W(T(ss_j))$ is the priority associated with the statement type $T(ss_j)$.
All suspicious statements are ranked by the updated suspicious scores in descending order, and developers can inspect them in order.

\section{Evaluation}
\label{section:Evaluation}
\subsection{Research Questions}
The evaluation investigates the following research questions:
\begin{itemize}[leftmargin=0.5cm,topsep=0.06cm]
	\item \textbf{RQ1:} Are different types of statements equally error-prone? If not, which kinds of statements are significantly more/less error-prone than others?
	\item \textbf{RQ2:} Does the proposed approach outperform the state-of-the-art fine-grained fault localization techniques? If yes, to what extent?
	\item \textbf{RQ3:} How does the performance of the proposed approach vary in locating different types of faulty statements?
	\item \textbf{RQ4:} Can we further improve the proposed approach by assigning priorities to all statements besides the special statements selected in Section~\ref{sub:priorities}?
	\item \textbf{RQ5:} Can we further improve the proposed approach by leveraging the median  instead of the average in Formula~\ref{qua:Wt}?	
\item \textbf{RQ6:} Should we apply multi-level sorting to fault localization by leveraging both SBFL scores and priorities of statement types?    	
    \item \textbf{RQ7:} Can we improve other SBFL algorithms in the same way as we improve $Ochiai$? If yes, to what extent? 	

\end{itemize}

\subsection{Dataset}
\label{sub:dataset}

The evaluation is based on the widely used Defects4J~\cite{D4j}. We select this dataset because it contains manually validated concise patches. It is critical that the patches are concise because concise patches are indispensable in the identification of faulty statements in subject programs (as explained in Section~\ref{sub:Statistics}), and the identified faulty statements should serve as the ground truth of the evaluation. Patches in some defect repositories, like iBUGS~\cite{dallmeier2007extraction} and ManyBugs~\cite{le2015manybugs}, are often not concise, i.e., they often contain bug-irrelevant changes. Consequently, bug-free statements modified for bug-irrelevant reasons (e.g., refactoring or requirements changes) could be incorrectly identified as faulty statements, which may result in inaccurate ground truth for the evaluation.

We exclude faulty versions whose patches do not remove or change any statements in the original faulty versions, i.e., such patches do nothing but insert new statements. We exclude such faulty versions because none of the involved statements are faulty according to the definition in Section~\ref{sub:Statistics}.
Notably, we also exclude such projects that contain less than 30 defects. As specified in Section~\ref{sub:Statistics}, the proposed approach identifies special statement types according to their relative error-proneness on different projects. If the involved projects contain only a small number of defects, the relative error-proneness on such projects has high degree of randomness, which could significantly influence the evaluation results.

The resulting data set is composed of seven faulty projects containing 376 defects. Details are presented on Table~\ref{D4j}.

\begin{table}[]
\renewcommand\arraystretch{1.1}
	\centering
	\caption{Dataset for Evaluation}
	\label{D4j}
	\begin{tabular}{|c|c|c|c|c|c|c|}
		\hline ID & Projects &  \#KLOC &\tabincell{c}{ \#Involved Bugs}\\ \hline
		Math&\tabincell{c}{Commons Math}& 85 &75 \\ \hline
		Closure&\tabincell{c}{Closure Compiler}&90 &83	 \\ \hline
		Lang&\tabincell{c}{Commons Lang}&22 &34	 \\ \hline
		Cli&\tabincell{c}{Commons Cli}&4&32	 \\ \hline
		JSoup&\tabincell{c}{Jsoup HTML parser}&14 &53	 \\ \hline
		Databind&\tabincell{c}{Jackson Databind}&71 &64	 \\ \hline
		Compress&\tabincell{c}{Commons Compress}&14 &35 \\ \hline
  \multicolumn{2}{|c|}{Total} & 300 &376 \\ \hline
	\end{tabular}
\end{table}

\begin{table*}[]	
\renewcommand\arraystretch{1.1}
	\centering
	\caption{Error-Proneness of Different Statements}
	\label{StatisticalResult}
	\begin{tabular}{|c|c|c|c|c|c|c|c|c|c|c|}
		\hline  & \tabincell{c}{Types of Statements}  &\tabincell{c}{\# Suspicious Statements}  &\tabincell{c}{Relative  Error-Proneness (RP)}  &Average of RP  &Median of RP\\ \hline		
		1 & Expression  &135,581 & [0.48, 1.22]&0.79&0.72\\ \hline		
        2 & Return  &75,887 &[0.39, 1.46]&0.89&0.91\\ \hline	
		3 & If &74,699  &[0.70, 2.62]&1.47&1.28\\ \hline		
		4 & Variable Declaration & 61,299& [0.63, 1.17]&0.96&1.02\\ \hline		
        5 & Constructor Invocation&7,959 &[0, 0] &0&0\\ \hline
        6 & For  &5,863 &[0, 1.37]&0.56&0.39\\ \hline	
		7 & Enhanced For&5,630  &[0, 2.46]   &1.02&1.28\\ \hline		
\end{tabular}
\end{table*}
\subsection{Experiment Design}
\subsubsection{RQ1: Assumption:} RQ1 concerns the assumption taken by \emph{HybridAFL}, i.e., different types of statements are not equally error-prone. To answer RQ1, we compute the average error-proneness for each type of statements on the selected faulty projects (as specified in Section~\ref{sub:dataset}). We also analyze the variation of the error-proneness among faulty projects by comparing their ranges, averages, and medians. To visualize the difference, we draw a box-plot to illustrate how often special statements are more/less error-prone than others.

\subsubsection{RQ2: Performance:}
\label{setting:RQ2}
RQ2 concerns the performance of \emph{HybridAFL}. To answer Rq2, we compare it against the widely used SBFL algorithm \emph{Ochiai}~\cite{Ochiai}. It is selected for comparison because of the following reasons. First, it represents the state of the art, and is widely employed in automated program repair~\cite{SimFix,PraPR}. Second, it serves as the baseline of \emph{HybridAFL}, and thus comparing it against \emph{HybridAFL} would reveal the benefit of leveraging statement types. 

We apply \emph{HybridAFL} and \emph{Ochiai} to the selected dataset independently, and compute their performance, i.e., \emph{absolute waste effort} (AWE)~\cite{PageRank}. 
 Given a faulty program and a ranked list of statements, AWE equals the ranking number (position) of the faulty statement.  
AWE is selected because it accurately represents the amount of effort wasted by automated program repair tools in generating invalid patches because of inaccurate fault localization.


Notably, the evaluation follows the widely used taken-one-out pattern. On each fold of the evaluation, a single project of the selected dataset is taken as testing data whereas other projects are taken as training data. Each of the projects is taken as testing data for once.

\subsubsection{RQ3: Strength and Weakness:}

RQ3 concerns the strength and weakness of \emph{HybridAFL}.  Because \emph{HybridAFL} gives different priorities  to different types of statements, it is likely that it is good at locating some kinds of statements (receiving highter priorities)  whereas poor at locating other statements. To answer this question, we classify faulty statements in the selected dataset according to statement types, and compute how efficient \emph{HybridAFL} is in locating such faulty statements.

\subsubsection{RQ4: Selection of Special Statement Types:} The key idea of the proposed approach is to assign special priorities to such types of statements that are steadily more/less error-prone than the average. To this end, in Section~\ref{sub:priorities} we identify such statement types according to their relative error-proneness, and compute priorities for each of them. Other statement types that are not selected receive the default priority (i.e.,1). To validate the necessity of the selection, we disable the selection, i.e., assigning priorities to all statement types according to Equation~\ref{qua:Wt}, and repeat the evaluation.

\subsubsection{RQ5: Median vs. Average:} RQ5 concerns how the error-proneness of a given statement type should be computed from a corpus of faulty projects. In Section~\ref{section:Approach} (Equation~\ref{qua:Wt}), we take the average (arithmetic mean) of the error-proneness on different projects as the final value. An alternative approach is to take the median value. To this end, we replace the average in Equation~\ref{qua:Wt} with the median, and repeat the evaluation as specified in Section~\ref{setting:RQ2}.

\subsubsection{RQ6: Multi-level Sorting:} RQ6 investigates whether multi-level sorting can further improve the performance of the proposed approach. Multi-level sorting~\cite{ProFL} is to rank items according to an attribute (called first attribute) first and then leverage additional attributes (called secondary attributes) to further rank such items that are given equivalent priorities according to the first attribute. In our case, we may rank suspicious statements by SBFL scores first and then by error-proneness of statement types, or vice versa. RQ6 would investigate whether such multi-level sorting is a good alternative to the hybrid approach in Equation~\ref{eq:NewScore}.

\subsubsection{RQ7: Effect on Other SBFL Algorithms:} RQ7 investigates whether the key idea of the proposed approach (i.e., leveraging statement types to enhance state-of-the-art fine-grained fault localization algorithms) can be applied successfully to other SBFL algorithms. To answer this question, we replace \emph{Ochiai}~\cite{Ochiai} in Equation~\ref{eq:SBFLO} with  \emph{Jaccard}~\cite{Jaccard}, $DStar$~\cite{DStar} and $Barinel$~\cite{Barinel}, and repeat the evaluation as specified in Section~\ref{setting:RQ2}. Such SBFL algorithms are selected because they are the most popular SBFL variants~\cite{APRSurvey,DStar}.


\subsection{Results and Analysis}
\label{sub:results}
\subsubsection{RQ1: Some Types of Statements are Significantly More Error-Prone Than Others:}  To answer RQ1, we compute the relative error-proneness (i.e., $RP(pr_i,t_j)$ in Equation~\ref{eq:EPscore}) for each of the statement types on each of the subject programs. Notably, the relative error-proneness $RP(pr_i,t_j)$ indicates to what extent statements of type $t_j$ are more/less error-prone than others in the same project. We do not present/compare the absolute error-proneness because it varies significantly among faulty projects, heavily dependent on the absolute numbers of suspicious statements in the projects rather than statement types.

\begin{figure}
	\begin{center}
		\includegraphics[width=0.5\textwidth]{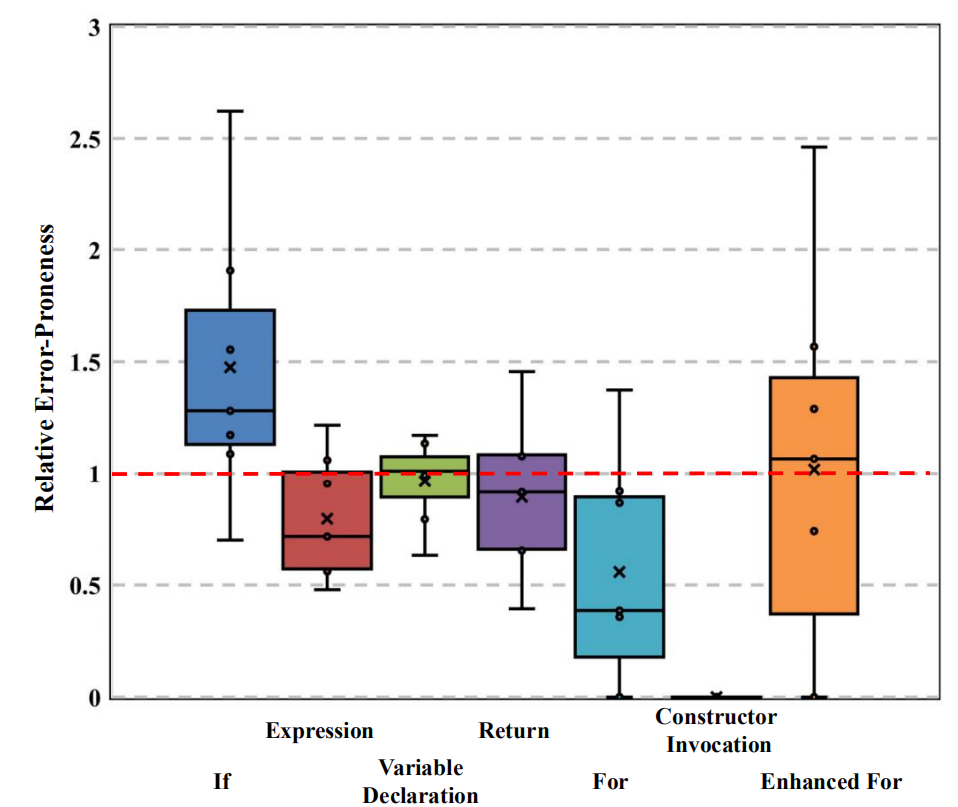}
		\caption{Relative Error-Proneness Varying among Programs}
		\label{fig:Distribution}
	\end{center}
\vspace{-0.4cm}
\end{figure}

Results of the analysis are presented on Table~\ref{StatisticalResult}. Notably, for space limitation Table~\ref{StatisticalResult} presents only popular statement types that account for at least 1\% of the suspicious statements in the subject programs. All of the statement types on Table~\ref{StatisticalResult} together account for more than 97\% of the suspicious statements. To visualize the different, we also present the relative error-proneness as a box-plot in Figure~\ref{fig:Distribution}. The horizontal axis on the figure represents different statement types whereas the vertical axis represents the relative error-proneness of different types of statements.  A box plot is constructed from five values: the minimum value, the first quartile, the median, the third quartile, and the maximum value. Notably, the \emph{enhanced for} statements (called \emph{EnhancedForStatement} in Eclipse JDT~\cite{JDT}) are such statements that follow the pattern `\emph{for (FormalParameter : Expression )}' where \emph{FormalParameter} is a single variable declaration without initializers. In contrast, the traditional \emph{for} statements (called \emph{ForStatement} in Eclipse JDT) follow pattern `\emph{for([ForInit];[Expression];[ForUpdate])}'. The red dashed line in Fig.~\ref{fig:Distribution} represents the average RP of all suspicious statements (i.e., RP=1.0) that serves as the baseline for the comparison.

\begin{table*}[]
\renewcommand\arraystretch{1.1}
	\centering
	\caption{Absolute Waste Effort (AWE)}
	\label{AWE}
	\begin{tabular}{|c|c|c|c|c|c|c|c|c|c|}
		\hline \diagbox{Approaches}{Projects}  & Math &Closure&Lang &Cli   &JSoup &Databind  &Compress&\textbf{Overall} \\\hline
	 Ochiai ($x$) & 7,227 &41,306 & 1,220 & 1,260 &3,468  &7,327  & 2,425& {64,223}\\\hline
	 HybridAFL ($y$) &5,524  &37,448 & 907  & 1,103 &3,786  &7,445  & 2,066&{58,279} \\\hline
	Relative Reduction ($1-y/x$) &23.6\%&9.3\% &25.7\%&12.5\%&-9.2\%& -1.6\%& 14.8\%&{{9.3\%}} \\	\hline
	\end{tabular}
\end{table*}
From Table~\ref{StatisticalResult} and Figure~\ref{fig:Distribution}, we make the following observations:
 \begin{itemize}[leftmargin=0.5cm,topsep=0.06cm]
 	\item First, some statement types are significantly more error-prone than others. For example, the average relative error-proneness (1.47) of \emph{if} statements  is significantly higher than that (0.56) of \emph{for} statements.
    \item Second, some statement types are steadily more error-prone than the average. For example, \emph{if} statement is more error-prone than the average on 6 out of the 7 projects. During fault localization, we should pay more attention to such kind of statements to speed fault localization.
    \item Third, some statement types are steadily less error-prone than the average. For example, \emph{constructor invocation} statements are never found faulty. Other statements, e.g., \emph{for} and \emph{expression} statements, are also steadily less error-prone than the average.
 \end{itemize}

We conclude based on the preceding analysis that different types of statements are not equally error-prone.  Some kinds of statements, e.g., \emph{if} statements, are significantly more error-prone than others whereas some kinds of statements, e.g., \emph{constructor invocation}, \emph{for}, and \emph{expression} statements, are significantly less error-prone than others. The conclusion validates the assumption that serves as the basis of the proposed approach.

\subsubsection{RQ2: HybridAFL Improves the State of the Art:}
To answer RQ2, we apply the proposed approach (HybridAFL) and traditional SBFL (Ochiai) to the selected faulty projects  independently. Results of the take-one-out (k-fold) evaluation are presented on Table~\ref{AWE}.  Absolute waste effort (AWE) represents how much effort has been wasted because of the inaccurate faulty localization. With regard to automated program repair (APR) that depends on the automated fault localization, AWE represents how many times APR tools have tried to fix a suspicious statement before they finally come to fix the faulty statement. Consequently, the smaller AWE is, the more efficient APR tools would be. The second row of Table~\ref{AWE} presents the performance (AWE) of Ochiai wheras the third row presents the performance of HybridAFL. The last row presents to what extent HybridAFL can reduce AWE with regard to the baseline approach Ochiai: positive percentages on the last row represent reduction in AWE whereas negative percentages represent increase in AWE.

From Table~\ref{AWE} we make the following observations:
\begin{itemize}[leftmargin=0.5cm,topsep=0.06cm]
\item First, HybridAFL significantly outperforms Ochiai.  On 5 out of the 7 projects, HybridAFL results in smaller AWE than that of Ochiai. The reduction in AWE varies from 9.3\% (on \emph{Closure}) to 25.7\% (on \emph{Lang}). On average, the reduction of AWE on all of the involved subject projects is 9.3\%.
\item Second, although the overall performance of HybridAFL is better than that of Ochiai, it is not guaranteed that HybridAFL can always outperform Ochiai. On 2 out of the 7 projects, i.e., \emph{JSoup} and \emph{Databind},  HybridAFL results in increased AWE. On \emph{Databind}, HybridAFL increases AWE slightly by only 1.5\%. However, on \emph{JSoup}, it increases AWE significantly by 9.2\%. The major reason for the increase is that the relative error-proneness of different statement types varies significantly from project to project. For example, on \emph{Jsoup}, the relative error-proneness (0.7) of \emph{if} statements  is significantly smaller than that (1.6) on other projects. As a result, the priorities of statement types learned from other projects fail to work on \emph{Jsoup}, which results in increased AWE.
\end{itemize}

We conclude based on the preceding analysis that the proposed approach frequently outperform the state-of-the-art approach, and it successfully reduces absolute waste effort by 9.3\% on average.

\begin{table*}[]
\renewcommand\arraystretch{1.1}
	\centering
	\caption{Effect on Different Types of Statements}
	\label{RQ3}
	\begin{tabular}{|c|c|c|c|c|c|c|c|c|c|}
		\hline   & If &\tabincell{c}{Variable\\ Declaration}  &Return  & Expression   &For &\tabincell{c}{Constructor \\ Invocation}  &Others   \\\hline
		Priorities &1.36    & 1   & 1  & 0.75  &0.4 &0 &1 \\\hline
		\#Improved Cases  ($x$)& 66 &15 & 20 & 1 &0 &0  &6  \\\hline
		\#Decreased Cases($y$)&0  &11 & 19  & 50 &3 &0 &0   \\\hline
		\#Draw Cases($z$)&3&4 &6&22&0&0& 5\\	\hline
		Chance of Improvement ($x/(x+y+z)$)&95.7\% &50.0\% &44.4\% &1.4\% &0\%&- &54.5\% \\ \hline
       Chance of Reduction ($y/(x+y+z)$)&0\% &36.7\% &42.2\% &68.5\% &100\%&- &45.5\%\\	\hline
	\end{tabular}
\end{table*}
\begin{table*}[]
\renewcommand\arraystretch{1.1}
	\centering
	\caption{Selection of Special Statement Types Reduces AWE}
	\label{DifProcessOrNot}
	\begin{tabular}{|c|c|c|c|c|c|c|c|c|c|}
		\hline \diagbox{Setting}{Programs}  & Math &Closure&Lang &Cli   &JSoup &Databind  &Compress&\textbf{Overall} \\\hline
		Selection Enabled ($x$) &5,524  &37,448 & 907  & 1,103 &3,786  &7,445  & 2,066&58,279 \\\hline
		Selection Disabled ($y$) &6,420&36,933 &1,140&1,108& 3,917 &8,240& 2,003&59,643 \\	\hline				
		Relative Reduction in AWE ($1-y/x$) &-16.2\%&1.4\% &-25.7\% &-0.5\% &-3.5\% &-10.7\% &3.0\% &-2.3\%
		\\	\hline
	\end{tabular}
\end{table*}
\subsubsection{RQ3: Good at Locating High-Priority Statements:}
To answer RQ3, we compute the performance of the proposed approach on each statement types. Evaluation results are presented on Table~\ref{RQ3}. Each column of the table presents the effect of the proposed approach in locating a specific type of faulty statements. The second row of the table presents the priorities assigned to different statement types (according to Equation~\ref{qua:Wt} in Section~\ref{section:Approach}).  The third row presents how often the proposed approach locates the faulty statements more quickly than the baseline approach (SBFL) whereas the fourth row presents how often it locates the faulty statements less quickly than the baseline. The fifth row presents how often they draw a tie.

From Table~\ref{RQ3}, we make the following observations:
\begin{itemize}[leftmargin=0.5cm,topsep=0.06cm]
\item The proposed approach works well in locating faulty \emph{if} statements. On 66 out of the 69 faulty \emph{if} statements, the proposed approach outperforms the baseline, i.e., the faulty statements are ranked in the front of the positions suggested by the baseline approach. We also notice that on average the proposed approach reduces  AWE by 32.7\% on average in locating such faulty \emph{if} statements.
\item The proposed approach has negative effect on less error-prone statement types, e.g., \emph{expression} and \emph{for} statements. Such statements are less error-prone, and thus are given lower priorities by the proposed approach. As a result, their suspicious scores are underestimated, and they are ranked in lower positions.
\item The proposed approach has minor positive effect on normal statement types that receive the default priority (i.e., 1). For example, on 50\% of the \emph{variable declaration} statements, the proposed approach outperforms the baseline whereas  it loses on only 36.7\% of the cases. As a result, the proposed approach reduce AWE by 2.8\% in locating \emph{variable declaration} statements. The reason for the minor positive effect is that numerous less error-prone statements (especially \emph{expression} statements) are ranked behind by the proposed approach, and thus such normal statements like \emph{variable declaration} receive higher ranking.
\end{itemize}

We conclude based on the preceding analysis that the proposed approach is good at locating more error-prone statement types whereas it could be less effective in locating less error-prone statement types. Notably,  more error-prone statement types account for a larger proportion of faulty statements. For example,  \emph{if} statements alone account for more than 20\% of the faulty statements in Defects4J~\cite{D4j}.

\subsubsection{RQ4: Selection of Special Statement Types:}
To answer RQ4, we assign special priorities to all statement types discovered in the subject programs according to Equation~\ref{qua:Wt}, and repeat the evaluation.
Evaluation results are presented on  Table~\ref{DifProcessOrNot}. The first column specifies the setting of the evaluation, i.e., whether we assign special priorities to selected statement types only (selection enabled) or assign priorities to all statement types (selection disabled). The other columns present the performance (AWE) of the proposed approach varying among different projects. The last row presents to what extent disabling the selection would decrease AWE .

From Table~\ref{DifProcessOrNot}, we observe that the selection of special statement types is useful. Disabling the selection results in decreased performance (i.e., increased AWE) on 5 out of the 7 projects, and the average AWE on all projects is increased by 2.3\%.  The reason is that the relative error-proneness of some statement types (e.g., \emph{variable declaration})  varies significantly among projects. For example, on four out of the seven projects \emph{variable declaration} is more error-prone than the average whereas it is less error-prone than the average on the other three projects. As a result, assigning a universal priority to \emph{variable declaration} could randomly result in negative/positive effect depending on the characters of involved faulty projects. Assigning priorities to special statement types only (that are steadily more/less error-prone than the average) helps to reduce the randomness, and thus improves the robustness of the proposed approach.

\begin{table*}[]
	\centering
	\caption{Derivation of Priorities Has Minor Effect on AWE}
	\label{DifAveVSMedian}
	\begin{tabular}{|c|c|c|c|c|c|c|c|c|c|}
		\hline \diagbox{Priorities}{Projects}  & Math &Closure&Lang &Cli   &JSoup &Databind  &Compress&\textbf{Overall} \\\hline
		Average ($x$) &5,524  &37,448 & 907  & 1,103 &3,786  &7,445  & 2,066&58,279 \\\hline		
		Median ($y$) &5,666&37,523 &986&1,131& 3,648 &7,585& 2,047&58,586 \\	\hline
		Relative Reduction in AWE ($1-y/x$) &-2.6\%&-0.2\% &-8.7\% &-2.5\% &3.6\% &-1.9\% &-0.9\% &-0.53\%
	\\	\hline
	\end{tabular}
\end{table*}
\begin{table*}[]
	\centering
	\caption{Multi-level Sorting Increases AWE}
	\label{DoubleLayer}
	\begin{tabular}{|c|c|c|c|c|c|c|c|c|c|}
		\hline \diagbox{Sorting}{Projects}  & Math &Closure&Lang &Cli   &JSoup &Databind  &Compress&\textbf{Overall} \\\hline		
		Hybrid  &5,524  &37,448 & 907  & 1,103 &3,786  &7,445  & 2,066&58,279 \\\hline			
        Multi-Level (SBFL First) &6,716&39,762&1,177&1,196&3,385&7,357&2,238&61,931 \\	\hline	
        SBFL Only & 7,227 &41,306 & 1,220 & 1,260 &3,468  &7,327  & 2,425& 64,223\\\hline
		Multi-Level (Priority First) &4,701&52,101&1,068&1,518&9,965&36,144&2,419&107,916 \\	\hline
		Priority Only &6,459 &77,752 & 1,267 & 1,965 &13,405  &53,571 & 2,938& 157,357\\\hline
	\end{tabular}
\end{table*}
\begin{table*}[]
	\centering
	\caption{Relative Improvement on Various SBFL Algorithms}
	\label{DifSBFL}
	\begin{tabular}{|c|c|c|c|c|c|c|c|c|c|}
		\hline \diagbox{SBFL Algorithms}{Projects}  & Math &Closure&Lang &Cli   &JSoup &Databind  &Compress&\textbf{Overall} \\\hline		
		Ochiai &23.6\%&9.3\% &25.7\%&12.5\%&-9.2\%& -1.6\%& 14.8\%&9.3\% \\	\hline
		Jaccard &14.3\%&9.9\% &5.9\%&8.5\%&-9.2\%& 4.6\%& 15.0\%&8.8 \% \\	\hline
		DStar &15.3\%&10.1\% &14.0\%&8.1\%&-9.2\%& 3.4\%& 12.2\%&9.0 \% \\	\hline
		Barinel &14.3\%&9.9\% &9.0\%&8.4\%&-9.2\%& 7.2\%& 15.1\%&9.2\% \\	\hline
	\end{tabular}
\end{table*}
\subsubsection{RQ5: Average VS. Median:}
To answer RQ5, we replace the average of $nor\_RP(t_j)$ in Equation~\ref{qua:Wt} with the median of $nor\_RP(t_j)$, and repeat the evaluation. Evaluation results are presented on  Table~\ref{DifAveVSMedian}. The first column specifies whether the average or medial of $nor\_RP(t_j)$ is employed in Equation~\ref{qua:Wt}.  The other columns present the performance (AWE) of the proposed approach. The last row presents to what extent replacing the average with median can improve the performance (i.e., reducing AWE).

From Table~\ref{DifAveVSMedian}, we observe that replacing the average with median cannot improve the performance. The replacement results in increased AWE on 6 out of the 7 projects, and the average AWE is also slightly increased. However, we also notice that the impact of the replacement is minor, and in most cases the performance changes slightly by no more than 3\%.

\subsubsection{RQ6: Multi-level Sorting Does Not Work:}
To answer RQ6, we replace the hybrid formula (Equation~\ref{eq:NewScore}) with multi-level sorting and repeat the evaluation. The results are presented on  Table~\ref{DoubleLayer}. The first column present how we sort suspicious statements, and other columns present the resulting performance (AWE). `Hybrid' is the default setting of the approach. `Multi-Level (SBFL First)' is to rank suspicious statements by SBFL scores first while priorities of statement types are taken as the secondary sorting attribute.

From Table~\ref{DoubleLayer}, we make the following observations:
\begin{itemize}[leftmargin=0.5cm,topsep=0.06cm]
  \item First, compared to the default hybrid sorting, multi-level sorting is less effective, resulting in increased AWE on most of the evaluated programs. AWE is increased on average by 6.3\% (SBFL first) and 85.2\% (priority first)
  \item Second, compared to approaches leveraging both SBFL scores and priorities of statement types, leveraging SBFL only or priorities only results in the lowest performance. Compared to the proposed approach, SBFL only and priority only increase AWE  by 10.2\% and 170\%, respectively.
\end{itemize}

The evaluation results may suggest that dynamic test case execution (SBFL suspicious scores) and static information (statement types) should not be considered separately.

\subsubsection{RQ7: Improving Various SBFL Algorithms:}
To answer RQ7, we replace the SBFL algorithm (Ochiai) in Equation~\ref{eq:SBFLO} with its variants (i.e., Jaccard~\cite{Jaccard}, DStar~\cite{DStar}, and Barinel[ref]), and repeat the evaluation. Evaluation results are presented on Table~\ref{DifSBFL}. This table presents to what extent the key idea (assigning priorities to different statement types) can improve various SBFL algorithms, i.e., to what extent the AWE could be reduced compared to the baselines (SBFL algorithms).  For example, the cell on the last column and the last row suggests that integrating the priorities (derived in Section~\ref{sub:priorities}) with Barinel can reduce AWE by 9\% on average compared to Barinel alone.

From Table~\ref{DifSBFL},  we make the following observations:
\begin{itemize}[leftmargin=0.5cm,topsep=0.06cm]
\item First, the key idea improves all of the evaluated SBFL algorithms with regard to their performance. The improvement (i.e., reduction in AWE) varies slightly from 8.8\% to 9.3\%.
\item Second, the improvement depends more on projects than on SBFL variants. For example, the idea results in the greatest reduction in performance on project \emph{JSoup}, regardless of the employed SBFL variants.
\end{itemize}

We conclude based on the preceding analysis that the proposed key idea, i.e., assigning different priorities to different statement types in fault localization, could be applied to various SBFL variants. The application could result in significant improvement in performance, regardless of the employed SBFL variants.


\subsection{Threats to Validity}
\label{sub:threats}
A threat to construct validity is that the labels of faulty statements in the evaluation dataset could be incorrect, which may result in inaccurate calculation of the performance of the evaluated approaches. Notably, the employed dataset, i.e., Defects4J, does not explicitly specify faulty statements. To this end, we identify faulty statements by analyzing patches provided by Defects4J: Statements deleted or modified by the patches are faulty statements. However, it could be inaccurate  when  the patches are incomplete or unconcise. They could also be inaccurate when there are multiple valid patches for the same faulty program: Faulty statements identified according to different patches could be inconsistent.
However, inaccurate identification of faulty statements could significantly influence the computation of the performance in fault localization.

A threat to external validity is that the involved dataset is small and thus the conclusions drawn on such dataset may not be generalized to other datasets (i.e., faulty programs to be fixed). The negative effect of the limited size is twofold. On one side, the limited number of faulty statements/programs leveraged to derive the priorities for different statement types makes the resulting priorities questionable with regard to its representativeness.  On the other side, the limited number of faulty statements/programs leveraged as testing data reduces the conclusions' generalizability: The randomness of the results increases significantly with the decrease in the size of the testing data. To reduce the threat, we leverage the largest public defect repository (Defects4J) with manually validated concise patches. Notably, we do not leverage artificial-fault datasets where faults are injected  manually or automatically by mutation tools. Such artificial faults could be significantly different from real faults, and thus they may have significantly different error probability.

\section{Discussion}
\label{section:Discussion}


\subsection{Simple, Intuitive, and Lightweight}
Simpleness is a big plus for the proposed approach. First, the approach is simple and intuitive, making it easy to understand, implement, and use. The proposed approach is essentially  a slight adaption to the widely used SBFL algorithms. The key idea for the adaption, i.e., different types of statements are not equally error-prone, is intuitive. Second, because of the simpleness, the proposed approach could be integrated into various existing fault localization algorithms without complex adaption. For example, our evaluation results in Section~\ref{section:Evaluation} suggest that the proposed approach significantly improves all of the evaluated SBFL variants. The only required adaption  is to time the resulting suspicious scores (from baseline approaches) by fixed priorities (Equation~\ref{eq:NewScore}). In future, we will try to integrate it into other AFL techniques, e.g., mutation-based techniques and state-based techniques. Finally, the proposed approach itself, excluding the involved baseline algorithm (SBFL), is lightweight. Although it is time-consuming to determine the priorities of different statement types by analyzing a large corpus, it could be done offline and it is done once and for all. The online adaption depends on the types of suspicious statements only, and thus could be done efficiently. It is a significant advantage of the proposed approach compared to  spectrum-based fault localization and mutation-based algorithms that require time/resource-consuming execution of test cases.

\subsection{Lack of Large Defect Repositories}
As specified in section~\ref{section:Approach}, the proposed approach relies on defect repositories to derive priorities for special statement types. However, it is challenging to construct large repositories of real defects from the industry as well as concise patches (or explicitly specified faulty statements). To the best of our knowledge, Defects4J~\cite{D4j} is the largest one with less than one thousand real defects collected from open-source applications as well as their corresponding patches. The limited size of the available qualified defects  may reduce the representativeness  of the resulting priorities assigned to various statement types, and thus limits the practicability of the proposed approach. Notably, generating artificial defects by mutation tools would not help. The error probability of such artificial defects may not follow the same model as real defects do, and thus they could not be employed to infer the error-proneness of real statements. Notably, even with the available repository (Defects4J) only, our approach significantly improves the state of the art. 

\subsection{Programming Language Specific Models}
Although the proposed approach is generic, priorities derived from one programming language may not be directly applicable to other programming languages because of the following reasons. First, different programming languages may have different types of statements. For example, $EnhancedForStatement$  are common in modern programming languages like \emph{Java}, but they are often illegal in traditional languages like \emph{C} and \emph{C++}. Consequently,  priorities learned from \emph{C} or \emph{C++} programs might not work for \emph{Java} programs. Second, even the same statement type may follow different error probability models in different programming languages. Consequently, priorities derived from the error probability models could be language-specific as well.

%

\section{Conclusions and Future Work}
\label{section:Conclusion}
Accurate fault localization is critical for program debugging and program repairing. However, it remains challenging to locate faulty statements accurately and automatically. Inaccurate fault localization often results in significant effort wasted in inspection of bug-free statements. To this end, in  this paper, we propose a simple and intuitive approach to fine-grained automated fault localization. The key idea of the proposed approach is that different types of statements are not equally error-prone, and thus statement types should be exploited by statement-level fault localization. To the best of our knowledge, the proposed approach is the first hybrid approach to fine-grained fault localization that leverages both dynamic test case execution and static statement types. Our evaluation results on Defects4J suggest that the proposed approach significantly outperforms the widely used fine-grained fault localization approaches. Our data and script are publicly available~\cite{github}.

In the future, it is interesting to construct large fault repositories and to derive error-proneness of different statement types based on the resulting repositories. As discussed in Section~\ref{section:Discussion}, existing repositories of real faults are rather small, which limits the practicability of the proposed approach. However, automated or semi-automated construction of large fault repositories remains challenging because of the following reasons. First, it is difficult to distinguish bug-fixing commits from other commits automatically. Second, it is difficult to automatically distinguish bug-fixing changes from other changes (e.g., refactoring) within the same bug-fixing commit. In future, it is also interesting to explore additional static or dynamic information that could be leveraged for statement-level fault localization. The proposed approach may inspire a series of  hybrid approaches to fine-grained fault location that leverage different dynamic/static information.

\bibliography{acmart.bib}

\end{document}